# Giant and Oscillatory Junction Magnetoresistance via RKKY-like Spin Coupling in Spin Gapless $Mn_2CoAl/SiO_2/p$-Si Heterostructures

**Nilay Maji*, Subham Mohanty, Pujarani Dehuri**

Department of Physics and Astronomy, National Institute of Technology Rourkela, Rourkela-769008, Odisha, India

***Author to whom correspondence should be addressed:** majin@nitrkl.ac.in

## Abstract

Here, we report spin-selective transport and exceptionally large positive junction magnetoresistance (JMR) in sputter-deposited $Mn_2CoAl$/native-$SiO_2$/$p$-Si heterostructures. Highly ordered inverse-Heusler $Mn_2CoAl$ thin films with near-ideal XA chemical ordering ($S \approx 0.97$) and a Curie temperature of ~590 K are realized using magnetron sputtering process. The spin-gapless semiconducting nature of $Mn_2CoAl$ is experimentally supported by a weakly temperature-dependent resistivity with a very small negative temperature coefficient of resistance (TCR $\approx - 4.2 \times 10^{-9}\ \Omega \cdot m \cdot K^{-1}$) and a nonsaturating linear magnetoresistance over a wide range of magnetic fields and temperatures. A giant positive JMR of ~825% at 10 K and ~134% at room temperature is observed despite the presence of only a single ferromagnetic electrode. Systematic variation of the $SiO_2$ tunnel barrier thickness reveals a reproducible oscillatory sign reversal of the JMR accompanied by a monotonic decay in magnitude. This behavior reflects thickness-dependent modulation of spin-selective tunneling mediated by phase-coherent interfacial carriers. It can be described phenomenologically by an RKKY-like functional form without invoking conventional metallic exchange interactions. These results identify $Mn_2CoAl$/native-$SiO_2$/$p$-Si heterostructures as robust and scalable platforms for room-temperature spin-selective transport, with potential applications in semiconductor-compatible spin filters, magnetic field sensors, and reconfigurable spintronic logic elements.

## 1. Introduction

Spin-dependent transport across ferromagnet/semiconductor interfaces is a central requirement for semiconductor spintronics, with direct implications for nonvolatile memory, magnetic sensing, and low-power logic technologies. However, the practical realization of such devices remains hindered by fundamental challenges, most notably the conductivity mismatch between metallic ferromagnets and semiconductors, interfacial spin depolarization, and defect-assisted transport processes that obscure intrinsic spin-selective effects.[1-3] These issues motivate the exploration of alternative magnetic materials and interface concepts that can sustain robust spin-dependent transport while remaining compatible with scalable semiconductor processing. Spin-gapless semiconductors (SGSs) have emerged as a particularly promising material class in this context. The SGS concept, first proposed by Wang,[4] is defined by a unique electronic structure in which one spin channel is gapless while the opposite spin channel possesses a finite band gap, enabling highly spin-polarized carriers to be excited with minimal energy cost. This electronic structure offers intrinsic advantages over both conventional ferromagnets and half-metallic systems by reducing conductivity mismatch and suppressing hot-carrier-induced interfacial spin scattering.[5-7] Among SGS candidates, inverse Heusler alloys are especially promising due to their tunable band structure, high Curie temperatures, and compatibility with thin-film growth techniques. The inverse Heusler compound $Mn_2CoAl$ has been experimentally established as a prototypical SGS by Ouardi *et al.*, who reported its characteristic transport signatures and robust ferromagnetism.[8] Subsequent studies have confirmed its SGS-like transport behavior in both bulk and thin-film forms, underscoring its suitability as a spin-selective electrode for semiconductor-based heterostructures.[8-14]

From a device integration perspective, silicon-based spintronic architectures require CMOS-compatible fabrication, thermally stable interfaces, and scalable tunnel barriers.[15-17] In this regard, native silicon oxide represents a technologically realistic tunnel barrier that is ubiquitous in silicon microelectronics. While native oxides are often regarded as detrimental to spin transport, several studies have demonstrated that they can support reliable spin-dependent tunneling and junction magnetoresistance (JMR) in ferromagnet/semiconductor heterostructures when the interfacial chemistry is well controlled. Positive JMR[18-23] provides a direct electrical signature of spin-selective tunneling and is particularly attractive for magnetic-field sensing and reconfigurable device concepts. Moreover, carrier-mediated exchange interactions, such as Ruderman–Kittel–Kasuya–Yosida (RKKY)–type oscillatory coupling,[24,25] are known to enable distance-dependent switching between parallel and antiparallel spin configurations in metallic and hybrid systems. Extending such oscillatory spin-coupling concepts to semiconductor tunnel junctions offers an additional, largely unexplored degree of freedom for spintronic device engineering. In this work, we investigate spin-dependent transport in $Mn_2CoAl$/native-$SiO_2$/*p*-Si heterostructures fabricated using an industry-compatible magnetron sputtering approach. The choice of $Mn_2CoAl$ is motivated by its experimentally verified spin-gapless semiconducting character, high Curie temperature, and favorable conductivity matching with silicon. Our measurements reveal robust spin-selective transport manifested as a large positive junction magnetoresistance, persisting up to room temperature despite the presence of only a single ferromagnetic layer. Furthermore, systematic variation of the tunnel barrier thickness leads to an oscillatory sign reversal of the JMR, consistent with an RKKY-like exchange coupling mediated by spin-polarized carriers in the semiconductor channel. These results demonstrate the emergence of magnetic tunnel junction-like behavior without the need for multiple ferromagnetic electrodes and establish a scalable route toward thickness-engineered spin filters, compact magnetic-field sensors, and reconfigurable silicon-compatible spintronic devices.

## 2. Experimental Section

The $Mn_2CoAl/SiO_2$/*p*-Si heterostructure was formed by dc magnetron sputtering of $Mn_2CoAl$ thin films onto boron-doped *p*-Si substrates while intentionally retaining the native $SiO_2$ layer as an ultrathin tunnel barrier. Details of device fabrication, experimental methods, structural, chemical, morphological, and magnetic characterizations, and supplementary transport analyses are provided in the Supporting Information.

## 3. Results and Discussion

### 3.1 Resistivity measurement

The spin-gapless semiconducting nature of the $Mn_2CoAl$ thin film was investigated through temperature-dependent electrical resistivity measurements carried out using the four-probe technique, as depicted in the inset of Fig. 1(a). The $Mn_2CoAl$ film was deposited on an electrically insulating $Al_2O_3$ (sapphire) substrate to eliminate any parasitic conduction pathways. The sample dimensions were $10 \times 5$ mm$^2$ with a film thickness of approximately 85 nm, and a probe spacing of 2 mm was employed during the measurements. Owing to the high resistivity of the sapphire substrate, the measured transport response originates exclusively from the $Mn_2CoAl$ layer, enabling a reliable assessment of its intrinsic electrical properties. The measurements were carried out during slow cooling from room temperature down to 15 K to ensure thermal equilibrium at each data point. Figure 1(a) presents the longitudinal resistivity ($\rho_{xx}$) as a function of temperature in the range of 15-300 K. The resistivity exhibits an almost linear decrease with increasing temperature, indicative of a semiconducting-like transport behavior. From the $\rho_{xx}$-T dependence, the temperature coefficient of resistance (TCR) for the MCA film was extracted to be $-4.2 \times 10^{-9}$ Ω m K$^{-1}$. Notably, this value is several orders of magnitude smaller than that of conventional semiconductors, such as silicon, which typically exhibits a TCR of approximately $-7 \times 10^{-2}$ Ω m K$^{-1}$.[26] Instead, the obtained TCR closely matches those reported earlier for established SGS materials, including $Mn_2CoAl$ ($-1.4 \times 10^{-9}$ Ω m K$^{-1}$)[8] and CoFeCrAl ($-5 \times 10^{-9}$ Ω m K$^{-1}$),[27] further supporting the SGS-like character of our MCA film. At room temperature, the electrical conductivity ($\sigma_{xx}$) of the MCA film is estimated to be $4.54 \times 10^3$ S cm$^{-1}$, which is comparable to the conductivity reported for the prototypical SGS $Mn_2CoAl$ ($\sigma_{xx} \approx 2.44 \times 10^3$ S cm$^{-1}$ at 300 K).[8] In contrast, the conventional half-metallic ferromagnet $Co_2MnSi$ exhibits a much higher room-temperature conductivity on the order of $10^5$ S cm$^{-1}$,[28] nearly two orders of magnitude larger than that of the MCA film. This pronounced difference further highlights the distinct transport characteristics of SGS systems compared to half-metallic ferromagnets.

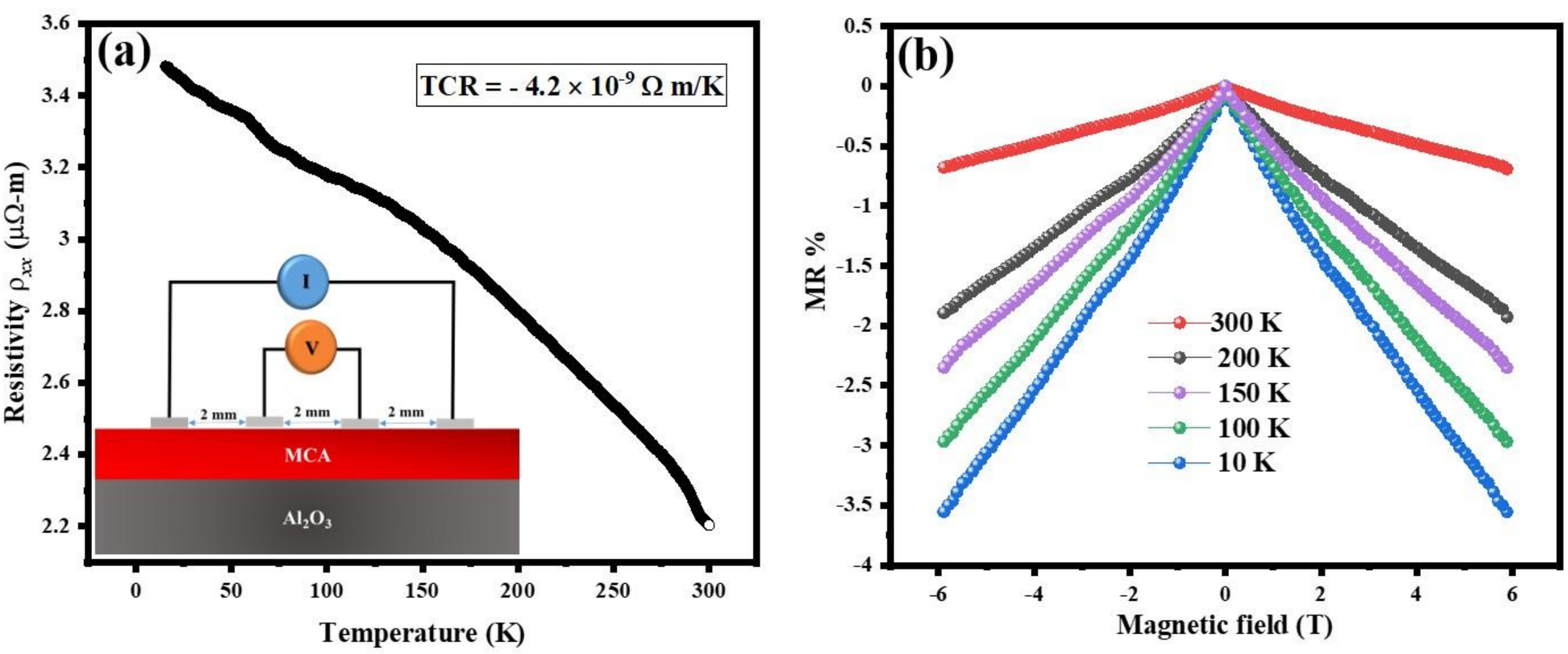

**Figure 1.** (a) Temperature-dependent longitudinal resistivity ($\rho_{xx}$) of a $Mn_2CoAl$ thin film measured on an insulating $Al_2O_3$ substrate using a four-probe geometry; the inset shows a schematic of the resistivity measurement configuration. (b) Magnetic-field-dependent magnetoresistance at 10–300 K showing a nearly linear, nonsaturating behavior up to ±6 T.

### 3.2 Magnetoresistance measurement

Magnetoresistance (MR) measurements were performed to further investigate the electronic transport properties of the $Mn_2CoAl$ thin film. As shown in Fig. 1(b), the MR recorded over the temperature range of 10-300 K exhibits a nearly linear dependence on the applied magnetic field and does not show any sign of saturation up to ±6 T. The magnitude of MR increases systematically with decreasing temperature, reaching its maximum value at 10 K, while remaining relatively weak at room temperature. The symmetric field dependence and absence of saturation across the entire temperature range are characteristic features of spin-gapless semiconductors and are consistent with previously reported quantum linear MR behavior in SGS systems[29]. These observations provide additional transport-based evidence supporting the spin-gapless nature of the $Mn_2CoAl$ thin film.

### 3.3 Transport property study

Next, we investigated the temperature-dependent electrical- and magneto- transport features across the $Mn_2CoAl/SiO_2/p$-Si heterostructure to study the dominant conduction mechanism through the native oxide barrier. Details of the temperature-dependent I–V characteristics, contact verification, transport modeling, and identification of the dominant tunneling mechanism are provided in the Supplementary Information.

#### *3.3.1 Magnetic field dependent I-V characteristics*

Figure 2(a) schematically illustrates the magnetotransport measurement geometry used for the $Mn_2CoAl/SiO_2/p$-Si heterostructure, where a constant dc bias is applied across the junction, and the resulting current is measured under an in-plane magnetic field. This current-perpendicular-to-plane configuration is highly sensitive to spin-dependent tunneling across the ultrathin $SiO_2$ barrier. Figures 2(b) and 2(c) present the current–voltage characteristics measured at 10 K and 300 K under zero field and an applied magnetic field of 6 T. At both temperatures, the magnetic field suppresses the junction current, most prominently in the forward-bias regime, while the reverse-bias current is comparatively less affected. This asymmetric field response indicates a magnetic-field-induced modulation of the tunneling probability of spin-polarized carriers, rather than changes in bulk silicon transport, consistent with the presence of positive junction magnetoresistance. The effect is more pronounced at 10 K, reflecting stronger spin selectivity and reduced thermal spin depolarization, yet remains clearly observable at room temperature, demonstrating robust spin-dependent transport across the heterojunction.

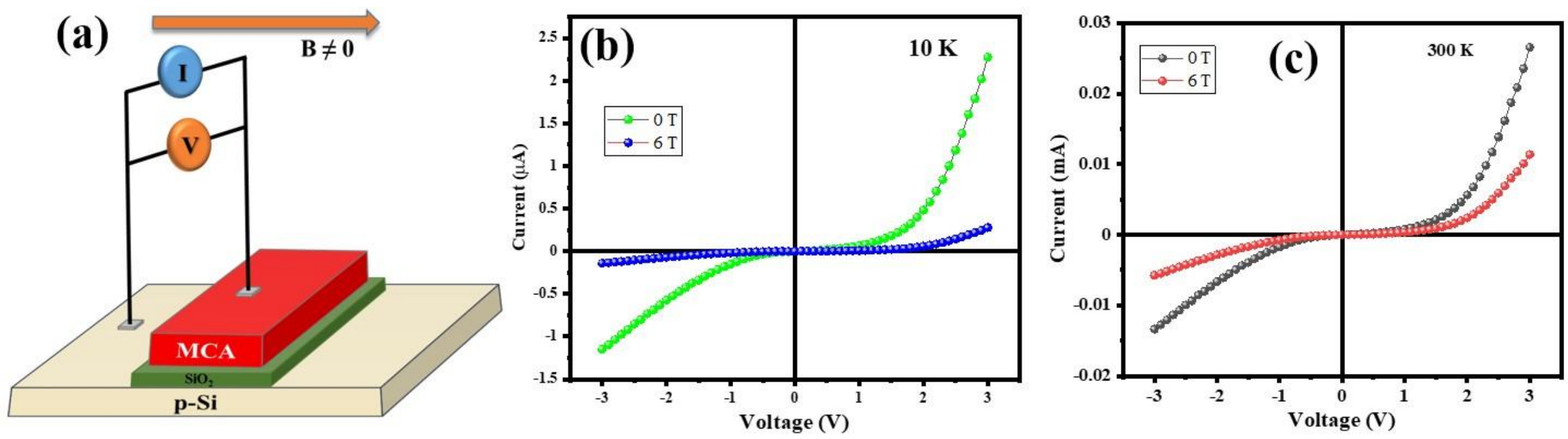

**Figure 2.** (a) Schematic of the measurement geometry used for magnetic-field-dependent *I–V* characterization. (b) *I–V* characteristics of the $MCA/SiO_2/p$-Si heterostructure measured at 10 K with and without applied magnetic field. (c) Corresponding *I–V* characteristics measured at 300 K. (d) Magnetic-field-dependent junction magnetoresistance (JMR) at 10 K and 300 K. (e) Schematic model of the $MCA/SiO_2/p$-Si heterojunction illustrating the formation of spin accumulation in *p*-Si in the absence of an external magnetic field. (f) Proposed schematic under an in-plane magnetic field of 6 T, showing enhanced spin-dependent scattering within the spin-accumulated region.

### *3.3.2 Magnetic field dependent Junction magnetoresistance (JMR)*

Figure 3(a) presents the magnetic-field-dependent junction magnetoresistance measured at 10 K and 300 K. The junction magnetoresistance is defined as

$$JMR = \frac{R(B) - R(0)}{R(0)} \tag{3}$$

where R(B) and R(0) are the junction resistances measured with and without an applied magnetic field, respectively. Robust positive junction magnetoresistance of approximately 825% is observed at 10 K, while a substantial positive junction magnetoresistance of about 134% is retained at room temperature. The positive sign of the junction magnetoresistance indicates that the resistance of the junction increases with increasing magnetic field, which is characteristic of spin-dependent tunneling across an insulating barrier.

The persistence of a large junction magnetoresistance at room temperature highlights the effectiveness of $Mn_2CoAl$ as a spin-gapless semiconductor injector and confirms that a significant degree of spin polarization is preserved during tunneling through the native $SiO_2$ barrier into p-type silicon.

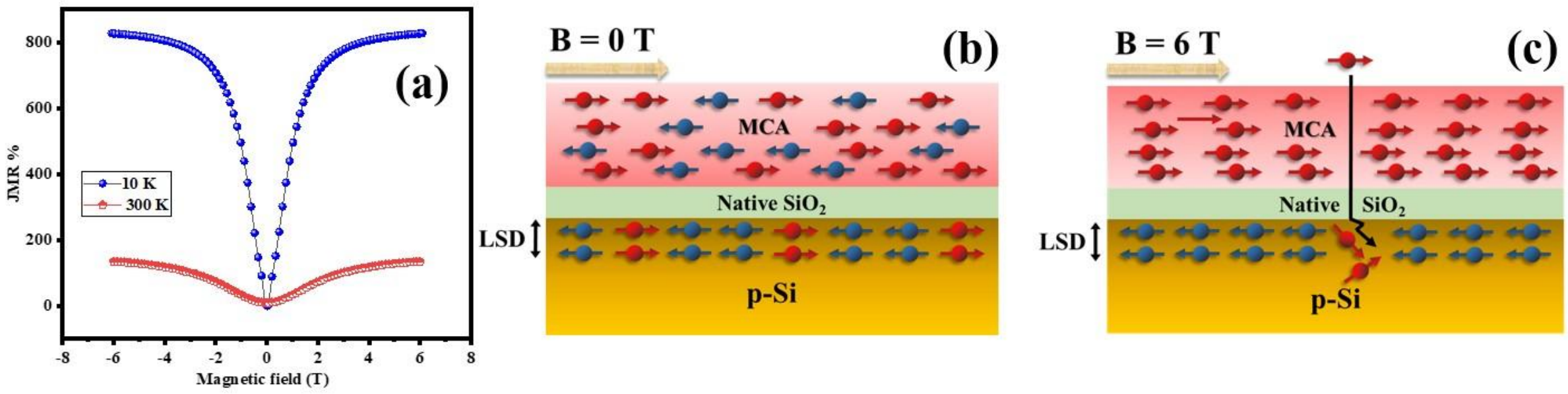

**Figure 3.** (a) Magnetic-field-dependent junction magnetoresistance (JMR) at 10 K and 300 K. (b) Schematic model of the MCA/$SiO_2$/*p*-Si heterojunction illustrating the formation of spin accumulation in *p*-Si in the absence of an external magnetic field. (c) Proposed schematic under an in-plane magnetic field of 6 T, showing enhanced spin-dependent scattering within the spin-accumulated region.

The emergence of a positive junction magnetoresistance (JMR) in the $Mn_2CoAl/SiO_2$/*p*-Si heterostructure is attributed to spin-dependent tunneling[30-32] across the ultrathin native $SiO_2$ barrier that mediates electronic coupling between the ferromagnetic $Mn_2CoAl$ layer and the *p*-Si channel. In the absence of an external magnetic field, bias-driven transport injects spin-polarized carriers from the spin-gapless $Mn_2CoAl$ into the *p*-Si via elastic tunneling, resulting in a localized spin accumulation near the $Si/SiO_2$ interface without a preferred spin orientation (Fig. 3(b)). Upon application of a sufficiently strong in-plane magnetic field, the $Mn_2CoAl$ magnetization approaches saturation through domain alignment, while the injected carriers in the adjacent *p*-Si acquire a field-stabilized spin polarization. The presence of the ultrathin insulating barrier renders the heterostructure functionally analogous to a magnetic tunnel junction, with $Mn_2CoAl$ acting as one spin-polarized electrode and the interfacial spin accumulation in *p*-Si forming an effective second spin-selective layer beneath the oxide. The experimentally observed positive JMR implies that the induced spin polarization in the semiconductor adopts an orientation effectively antiparallel to the $Mn_2CoAl$ magnetization, analogous to an antiparallel configuration in conventional magnetic tunnel junctions. This antiparallel alignment enhances spin-dependent scattering of tunneling carriers, leading to an increased junction resistance under magnetic field (Fig. 3(c)). We emphasize that this mechanism does not require long-range magnetic ordering in silicon, but instead arises from a localized, field-stabilized spin polarization confined to the vicinity of the $Si/SiO_2$ interface that directly participates in the tunneling process. With increasing temperature, carriers localized in the *p*-Si channel acquire sufficient thermal energy through phonon-mediated processes. This thermal activation destabilizes the induced spin-polarized regions near the interface and weakens the antiparallel spin alignment. Consequently, spin-dependent scattering is reduced, and the magnitude of the positive junction magnetoresistance decreases with increasing temperature.

### *3.3.3 Barrier layer thickness dependent JMR*

To investigate the interfacial origin of the junction magnetoresistance and to probe the role of the tunnel barrier in governing spin-dependent transport, we systematically investigated the dependence of the JMR on the thickness of the native $SiO_2$ layer. The oxide thickness was intentionally increased from approximately 2.0 nm to 2.5, 3.0, 3.5, and 4.0 nm while keeping all other deposition and measurement parameters unchanged at 300 K. Such controlled variation enables isolation of thickness-driven effects from changes in materials quality or device geometry. The persistence of polarized spin accumulation for $SiO_2$ barriers up to 4 nm arises from the spin-gapless electronic structure of $Mn_2CoAl$, in which a gapless majority-spin channel at the Fermi level coexists with a gapped minority-spin channel. This spin-resolved band asymmetry enforces single-spin-channel transport, suppresses interband spin-flip scattering, and preserves the spin polarization of tunneling carriers. Together with the moderate conductivity of $Mn_2CoAl$, which mitigates conductivity mismatch at the $Mn_2CoAl/SiO_2/p$-Si interface, these factors stabilize interfacial spin accumulation across comparatively thick oxide barriers.

Figure 4(a) shows the junction magnetoresistance (JMR) measured at 300 K as a function of applied magnetic field for $Mn_2CoAl/SiO_2/p$-Si heterostructures with different $SiO_2$ barrier thicknesses. As expected for tunneling-dominated transport, the absolute magnitude of the JMR decreases monotonically with increasing oxide thickness, reflecting the exponential suppression of spin-dependent tunneling probability. Remarkably, however, the *sign* of the JMR alternates systematically with increasing barrier thickness, switching between positive and negative values. This oscillatory sign reversal is reproducible across multiple devices and cannot be explained by monotonic variations in barrier height, tunneling probability, or interface resistance, all of which would yield a strictly decaying JMR without polarity inversion.

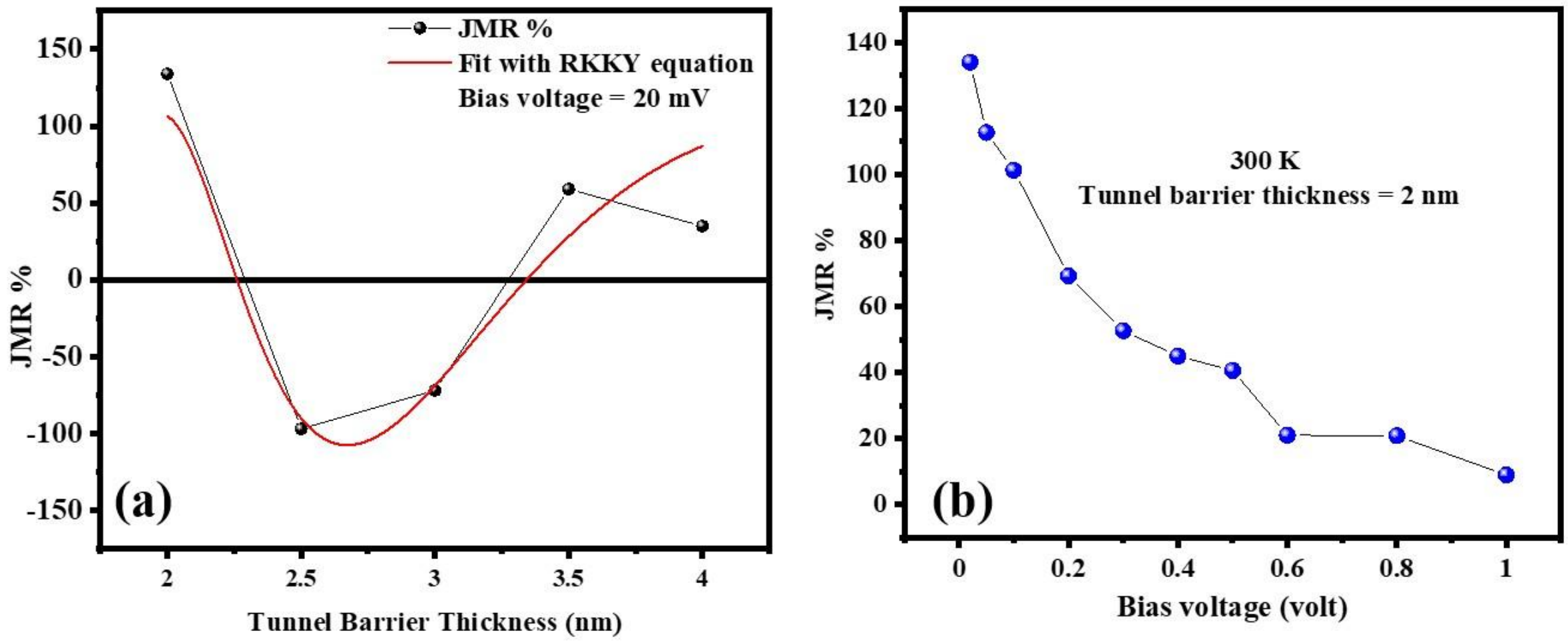

**Figure 4.** (a) Magnetic-field-dependent junction magnetoresistance (JMR) measured at 300 K for $Mn_2CoAl/SiO_2/p$-Si heterostructures with different $SiO_2$ barrier thicknesses. (b) Bias-voltage dependence of the JMR measured at 300 K for a representative device with fixed $SiO_2$ thickness.

The systematic oscillatory sign reversal of the junction magnetoresistance with increasing $SiO_2$ thickness reveals a distance-dependent modulation of the effective spin-selective tunneling configuration. Such oscillatory behavior is a generic hallmark of phase-coherent, carrier-mediated spin coupling across spatially separated regions and does not require the presence of a bulk metallic Fermi surface. In the present $Mn_2CoAl/SiO_2/p$-Si heterostructure, the relevant electronic states are confined near the $Si/SiO_2$ interface and participate in tunneling through an ultrathin oxide barrier. The observed oscillations therefore arise from thickness-dependent phase accumulation of spin-polarized tunneling carriers and are appropriately described using a Ruderman–Kittel–Kasuya–Yosida (RKKY)-like phenomenological oscillatory form, without invoking conventional metallic RKKY exchange. Here, the oscillatory functional dependence reflects interfacial, quasi-two-dimensional carrier mediation rather than bulk itinerant-electron coupling. We emphasize that the observed oscillations reflect phase-coherent modulation of spin-selective tunneling at the interface, rather than a true long-range magnetic exchange interaction.

Within this phenomenological framework, the effective interfacial coupling may be expressed in an asymptotic form analogous to oscillatory carrier-mediated interactions,

$$J(t) \propto \frac{\cos(2k_{eff}t + \phi)}{t^{n}}, \tag{4}$$

where t is the effective tunnel-barrier thickness, $k_{eff}$ represents an effective momentum scale associated with spin-polarized carriers confined near the $Si/SiO_2$ interface, $\phi$ is a phase factor determined by interfacial scattering, and $n$ reflects the dimensionality and spatial confinement of the mediating electronic states. Here, $k_{eff}$ captures the characteristic phase accumulation of interfacial carriers rather than a bulk Fermi wave vector.

Fitting the decay envelope of the JMR magnitude as a function of oxide thickness yields an exponent-$n \approx 1.9$. This value is consistent with coupling mediated by quasi-two-dimensional interfacial electronic states and is compatible with a scenario in which spin-polarized carriers are confined within a narrow region of *p*-Si near the $SiO_2$ interface. While this agreement is suggestive, the extracted exponent primarily serves to reinforce the interfacial and carrier-mediated nature of the observed oscillatory coupling.

#### *3.3.4 Bias voltage dependent JMR*

We have investigated the bias-voltage dependence of the junction magnetoresistance (JMR) for a fixed $SiO_2$ barrier thickness (2 nm), as shown in Fig. 4(b). The magnitude of the JMR exhibits a pronounced bias dependence, attaining its maximum value at low bias. The reason is attributed to the fact that the spin accumulation at the $Si/SiO_2$ interface remains robust as elastic tunneling dominates at low bias voltage. With increasing bias, the JMR amplitude decreases monotonically, which we attribute to the progressive activation of inelastic and field-assisted transport channels in the oxide, leading to a reduction of the effective spin polarization of tunneling carriers. The dominance of elastic tunneling at low bias voltage is consistent with our Poole–Frenkel analysis of the transport mechanism. At low applied bias, the electric field across the $SiO_2$ barrier is insufficient to activate field-assisted emission from localized defect states, as evidenced by the absence of linear behavior in the Poole–Frenkel representation in this regime. Consequently, charge transport occurs predominantly via direct, elastic tunneling near the Fermi level, where spin polarization is preserved most efficiently, leading to the maximum observed junction magnetoresistance. With increasing bias voltage, the enhanced electric field enables Poole–Frenkel-type conduction[21,33,34] through localized states in the oxide, introducing inelastic transport channels and resulting in a gradual suppression of the JMR amplitude. This clear crossover from elastic tunneling at low bias to field-assisted transport at higher bias directly correlates with the observed bias dependence of the magnetoresistance. Importantly, despite the strong bias dependence of the JMR magnitude, the polarity of the JMR remains unchanged over the entire bias range studied. This behavior indicates that the sign of the JMR is governed by the exchange-selected spin alignment across the tunnel barrier, which is fixed by the oxide thickness and is insensitive to bias-induced changes in the transport mechanism. These results further support the interpretation that bias primarily modulates the efficiency of spin-dependent tunneling, while the underlying exchange-mediated spin configuration remains intact.

## 4. Conclusion

In conclusion, the observation of large, room-temperature junction magnetoresistance and its systematic oscillatory modulation with tunnel barrier thickness demonstrates a robust interfacial spin-selective tunneling phenomenon in $Mn_2CoAl/SiO_2/p$-Si heterostructures. The results reveal a carrier-mediated, phase-coherent modulation of spin-selective tunneling at the interface, enabling oscillatory junction magnetoresistance in a semiconductor heterostructure without requiring multiple magnetic electrodes or metallic spacers. These findings provide a practical route toward thickness-engineered spin functionality in silicon-based spintronic devices.

## Acknowledgment

The authors acknowledge the Central Research Facility (CRF), National Institute of Technology Rourkela, for access to experimental characterization facilities. The corresponding author (N.M.) sincerely acknowledges financial support from the Seed Grant program of the National Institute of Technology Rourkela.

## AUTHOR DECLARATIONS

### Conflict of Interest

The authors have no conflicts to disclose.

**Nilay Maji:** Conceptualization (lead); Data curation (lead); Formal analysis (lead); Methodology (lead); Resources (lead); Software (lead); Validation (lead); Visualization (lead); Writing – original draft (lead); Writing – review & editing (equal). Project administration; Project Supervision. **Subham Mohanty:** Formal analysis (equal); Software (equal). **Pujarani Dehuri:** Formal analysis (equal); Software (equal).

## DATA AVAILABILITY

The data that support the findings of this study are available from the corresponding author upon reasonable request.

**Supporting Information**

**S1. Device fabrication**

$Mn_2CoAl/SiO_2/p$-Si heterostructures were fabricated using a controlled magnetron sputtering deposition process designed to preserve the native oxide tunneling layer while ensuring high crystalline and magnetic quality of the spin injector. Commercial boron-doped *p*-type Si (100) wafers with a hole concentration of approximately $4.5\times10^{16}$ $cm^{-3}$, a room-temperature resistivity of ~3 Ω·cm, and a thickness of 800 μm were used as substrates. The wafers were diced into rectangular pieces of typical dimensions 10 mm × 10 mm using a precision diamond cutter.

Prior to film deposition, the substrates were cleaned using a standard solvent-cleaning protocol to remove organic residues and surface particulates. The cleaning procedure involved sequential ultrasonic agitation in acetone, isopropyl alcohol, and deionized water for 10 min each, followed by drying under a high-purity nitrogen flow. No chemical oxide etching was performed in order to intentionally retain the native $SiO_2$ layer on the Si surface. This native oxide layer, with a thickness of approximately 2 nm as confirmed later by cross-sectional transmission electron microscopy, served as the tunnel barrier in the final heterostructure.

Thin films of the spin-gapless semiconductor $Mn_2CoAl$ were deposited on the cleaned *p*-Si substrates by dc magnetron sputtering using a stoichiometric $Mn_2CoAl$ alloy target (purity ≥ 99.9%). The sputtering chamber was evacuated to a base pressure better than $2.5 \times 10^{-7}$ Torr prior to deposition. High-purity argon (99.999%) was used as the sputtering gas, and the working pressure during deposition was maintained at 2.5 mTorr. The dc sputtering power was fixed at 50 W, yielding a deposition rate of approximately 1.0 nm/min. All depositions were carried out at room temperature without external substrate heating, and the film thickness was controlled by deposition time to achieve a nominal thickness of 85 nm. During deposition, the substrates were continuously rotated to ensure uniform film thickness and compositional homogeneity.

Following deposition, the samples were subjected to a post-deposition annealing treatment at 325 °C for 30 min in a controlled atmosphere to promote atomic ordering and stabilize the cubic phase of $Mn_2CoAl$. After annealing, the samples were allowed to cool naturally to room temperature before further characterization. This fabrication route results in well-defined $Mn_2CoAl$/native-$SiO_2$/*p*-Si heterostructures with sharp interfaces, enabling reliable investigation of spin-dependent transport and spin injection at room temperature.

**S2. Device characterization and measurement setup**

The structural properties of the $Mn_2CoAl$ thin films were examined using X-ray diffraction (XRD) performed on a Bruker D8 ADVANCE diffractometer (Bruker AXS, DAVINCI design) equipped with a Cu $K_\alpha$ radiation source (λ = 1.542 Å). XRD patterns were recorded at room temperature in the Bragg angle (2θ) range of $20° < 2\theta \leq 100°$, using a step size of 0.01° and a slow scan rate to accurately resolve the diffraction peaks and assess phase formation and crystallinity. The chemical composition and elemental states of the as-grown $Mn_2CoAl$ thin films were analyzed by X-ray photoelectron spectroscopy (XPS) using a PHI 5000 VersaProbe II scanning system. High-resolution core-level spectra were recorded to confirm chemical states and elemental stoichiometry. Surface morphology and topography of the $Mn_2CoAl$ thin films were investigated by atomic force microscopy (AFM) operated in tapping mode using an Agilent 5500 AFM system (Agilent Technologies). The AFM measurements provided quantitative information on surface roughness and film uniformity. The interfacial microstructure and thickness of the $Mn_2CoAl/SiO_2/p$-Si heterojunction were examined using cross-sectional high-resolution transmission electron microscopy (HRTEM) carried out on a JEOL JEM-2100 microscope (Japan). The HRTEM analysis enabled direct visualization of the native $SiO_2$ tunnel barrier and the structural quality of the interface. Magnetic properties of the $Mn_2CoAl$ thin films were characterized by measuring the magnetization (M) as a function of applied magnetic field (H) using a vibrating sample magnetometer (VSM, Lake Shore, Model 74034) at room temperature. The temperature-dependent electronic transport properties of the heterojunction were measured using a Keithley 2612 source measure unit (SMU) with a voltage resolution of 1 μV, in conjunction with an 8½-digit digital multimeter (Keithley 2000). Precise temperature control during electrical measurements was achieved using a Lake Shore 340 temperature controller (USA). This measurement configuration enabled reliable current-voltage, magnetotransport, and temperature-dependent transport characterization of the device.

## S3. Structural characterization

Figure S1(a) shows the X-ray diffraction (XRD) pattern of the $Mn_2CoAl$ thin film deposited on the *p*-Si (100) substrate. All observed diffraction peaks are indexed to a cubic inverse Heusler (XA) crystal structure, confirming the formation of single-phase $Mn_2CoAl$ without detectable secondary phases. The lattice parameter was calculated using Bragg's law, yielding a lattice constant of $a$ = 5.746 Å, which is consistent with reported values for $Mn_2CoAl$[1,2] in the XA phase.

Chemical ordering in inverse Heusler alloys is sensitively reflected by the presence of superlattice reflections. In the present case, the clear observation of both (111) and (200) reflections (as shown in the inset of Fig. S1(a)), in addition to the fundamental (220) peak, indicates a high degree of atomic ordering. For inverse Heusler alloys, the absence of both (111) and (200) peaks corresponds to a fully disordered A2 structure, while the presence of only the (200) peak is characteristic of B2-type disorder. The simultaneous presence of both superlattice peaks therefore confirms that the $Mn_2CoAl$ film is close to the ordered XA limit, with minimal antisite disorder. A quantitative estimate of the long-range chemical order was obtained from the intensity ratios of the superlattice and fundamental reflections. The order parameters are defined as

$$S^2=[I(200)/I(220)]_{theo}/[I(200)/I(220)]_{exp} \quad (1)$$

and

$$S^2(1-2\alpha)^2=[I(111)/I(220)]_{theo}/[I(111)/I(220)]_{exp}, \quad (2)$$

where S represents the long-range order parameter, α denotes the antisite disorder parameter, and I(hkl) denotes the integrated X-ray diffraction intensity of the Bragg reflection corresponding to the crystallographic plane with Miller indices (hkl). The extracted ratios indicate a high degree of chemical ordering with $\alpha \sim 0.01$ and $S \sim 0.97$, consistent with an ordered XA structure. The minor deviation from the ideal limit ($\alpha \sim 0$ and $S \sim 1$) is attributed to residual disorder below the detection limit of conventional XRD.

The average crystallite size D was estimated from peak broadening using the Debye–Scherrer equation,[3-5]

$$D = K\lambda/\beta\cos\theta, \quad (3)$$

where K=0.9 is the shape factor, λ=1.542 Å is the Cu $K_\alpha$ wavelength, and β is the full width at half maximum (FWHM) of the diffraction peak corrected for instrumental broadening. The calculated average crystallite size is $D \approx 25.13$ nm. The corresponding dislocation density δ,[6] which provides a measure of crystalline defect density, was estimated using

$$\delta = 1/D^2 \quad (4)$$

yielding δ=1.583×10$^{-3}$ nm$^{-2}$, indicative of relatively low defect density for a sputter-deposited inverse Heusler thin film.

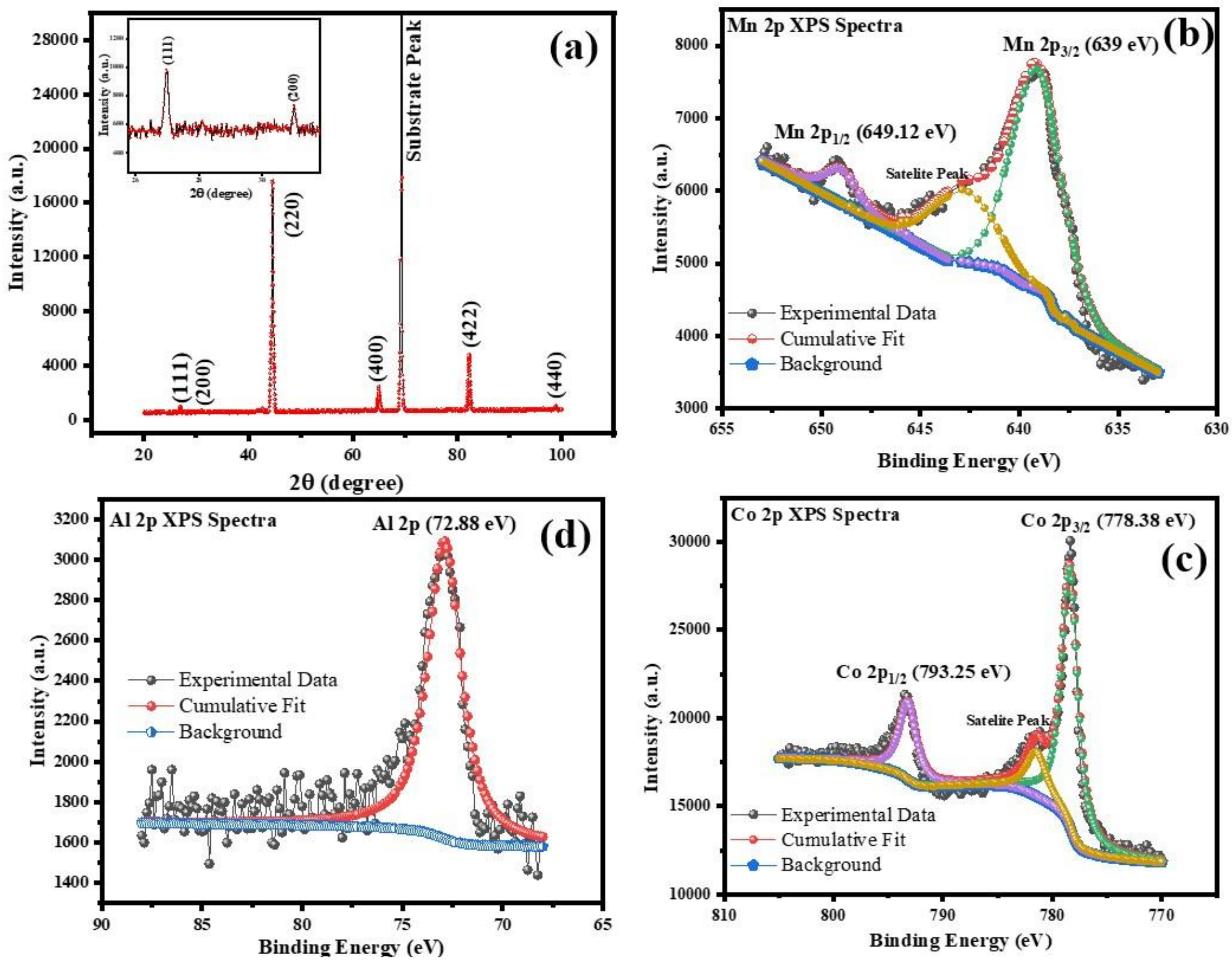

**Figure S1.** (a) High-resolution X-ray diffraction (HRXRD) pattern of the MCA film deposited on *p*-Si substrate, with the inset highlighting the (111) and (200) reflections indicative of a well-ordered crystal structure. (b–d) X-ray photoelectron spectroscopy (XPS) core-level spectra of the MCA thin film corresponding to the Mn 2p, Co 2p, and Al 2p states, respectively.

## S4. Chemical and electronic state analysis

The chemical composition and electronic states of the $Mn_2CoAl$ thin film were examined by X-ray photoelectron spectroscopy (XPS). Prior to XPS measurements, the $Mn_2CoAl$ thin film surface was gently etched using low-energy $Ar^+$ ion sputtering to remove adventitious surface oxidation and contaminants. Figure S1(b) displays the high-resolution Mn 2p spectrum. The Mn $2p_{3/2}$ and Mn $2p_{1/2}$ core-level peaks are centered at binding energies of 639.0 eV and 649.12 eV, respectively, with a spin–orbit splitting of ~10.1 eV. A weak satellite feature is observed at higher binding energy, indicating partial electron correlation effects typical of Mn-based Heusler alloys. The overall line shape and peak positions are consistent with Mn in a metallic bonding environment rather than a fully oxidized $Mn^{2+}$ or $Mn^{3+}$ state. The Co 2p core-level spectrum, shown in Fig. S1(c), exhibits prominent Co $2p_{3/2}$ and Co $2p_{1/2}$ peaks located at 778.38 eV and 793.25 eV, respectively, yielding a spin–orbit splitting of ~14.9 eV. The relatively weak satellite structure suggests that Co predominantly exists in an itinerant metallic state with strong hybridization of Co 3d states with neighboring Mn and Al atoms, which is characteristic of inverse Heusler compounds. Figure S1(d) presents the Al 2p spectrum, where a dominant peak is observed at 72.88 eV, corresponding to metallic Al bonded within the $Mn_2CoAl$ lattice. The absence of a pronounced higher-binding-energy Al–O component indicates that oxidation of Al is minimal and confined to the surface.

Overall, the XPS results confirm the correct elemental composition and predominantly metallic bonding character of Mn, Co, and Al in the $Mn_2CoAl$ thin film.[7] The observed chemical states are consistent with a chemically ordered inverse Heusler (XA) structure and support the structural conclusions drawn from XRD. Importantly, the preservation of metallic bonding and limited surface oxidation provides a favorable chemical environment for efficient spin injection and the enhanced room-temperature spin lifetime observed in the $Mn_2CoAl/SiO_2/p$-Si heterostructure.

**S5. Surface morphology analysis**

The surface morphology and topography of the $Mn_2CoAl$ thin film were examined using tapping-mode atomic force microscopy (AFM) to assess surface uniformity and roughness, which are critical for spin-dependent tunneling across the $Mn_2CoAl/SiO_2/p$-Si heterostructure. Figure S2(a) displays a representative three-dimensional AFM topographic image acquired over a scan area of $1 \times 1$ $\mu m^2$. The surface exhibits a dense, continuous, and homogeneous morphology without visible cracks, pinholes, or large particulates, indicating uniform film growth over the substrate.

Quantitative roughness parameters extracted from the AFM data, following ISO 25178 standards, are summarized in Fig. S2(b). The root-mean-square surface roughness (Sq) and arithmetic mean roughness (Sa) are determined to be 0.312 nm and 0.247 nm, respectively, confirming an atomically smooth surface. Figure S2(c) shows a representative line profile extracted from the AFM image, which provides an estimate of the film thickness. The measured thickness is approximately 81 nm**,** in close agreement with the nominal thickness of ~85 nm targeted during sputtering deposition by optimizing the deposition rate and growth time. Minor deviations are attributed to local surface undulations inherent to sputter-grown films.

**S6. Microstructural analysis**

Figure S2(d) shows a cross-sectional HRTEM image of the $Mn_2CoAl/SiO_2/p$-Si heterostructure, clearly revealing the layered configuration of the device. A uniform amorphous interfacial layer is observed between the crystalline $Mn_2CoAl$ film and the p-Si substrate. The thickness of this interfacial layer is estimated to be approximately 2 nm, consistent with that of native $SiO_2$. The absence of lattice fringes within this layer confirms its amorphous oxide nature. The $Mn_2CoAl$ layer exhibits clear lattice fringes, indicating good crystallinity, while the *p*-Si substrate retains its crystalline integrity near the interface. No evidence of interfacial reaction layers, pinholes, or significant interdiffusion is observed across the $Mn_2CoAl/SiO_2$ and $SiO_2/p$-Si interfaces, suggesting sharp and chemically abrupt interfaces even after post-deposition annealing. The presence of a continuous and uniform native $SiO_2$ tunnel barrier with well-defined interfaces is crucial for suppressing direct metallic contact and enabling spin-dependent tunneling. These interfacial characteristics provide strong microstructural support for the observed tunneling-dominated transport, large junction magnetoresistance, and enhanced room-temperature spin accumulation in the $Mn_2CoAl/SiO_2/p$-Si heterostructure.

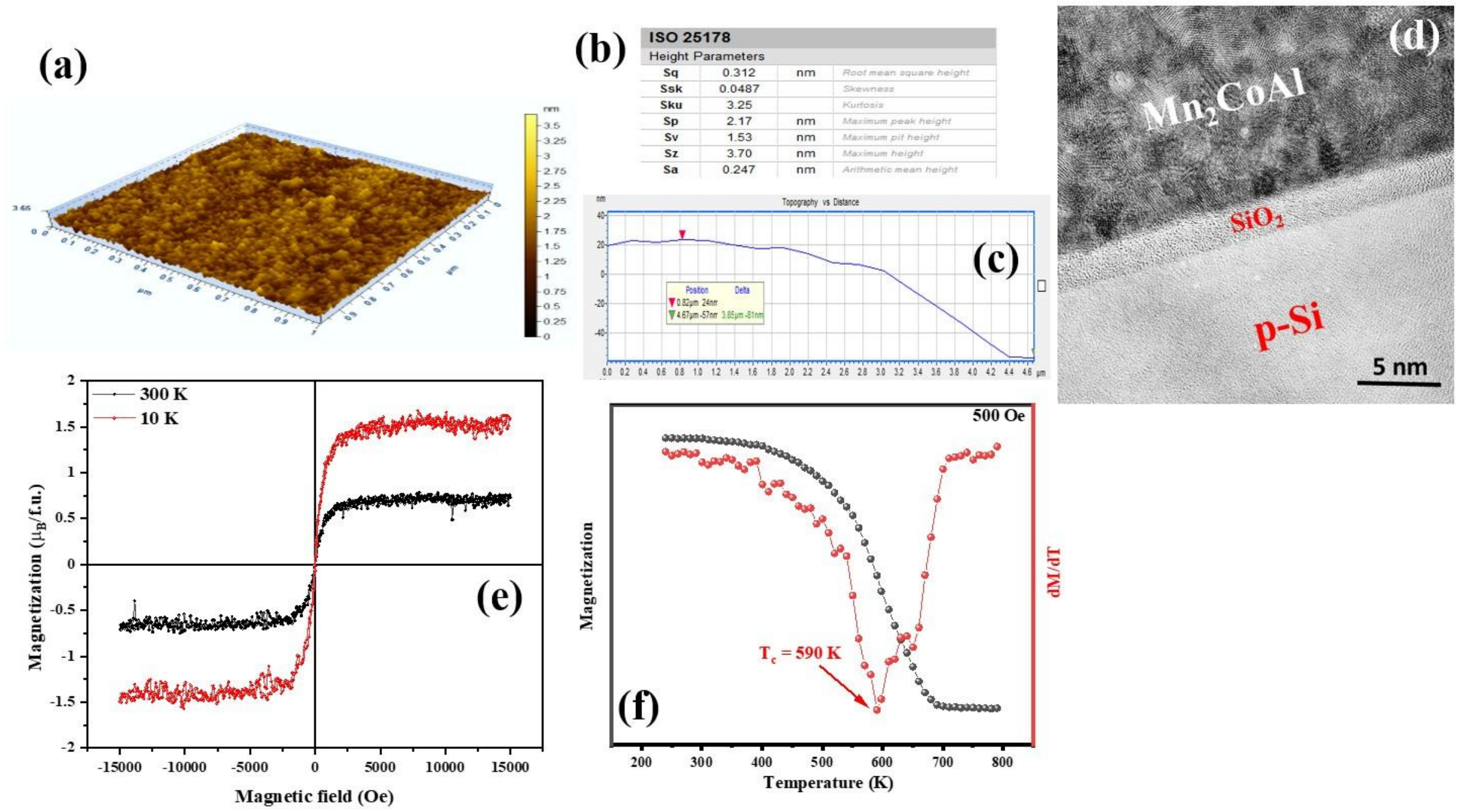

**Figure S2.** (a) Three-dimensional atomic force microscopy (AFM) surface morphology of the $Mn_2CoAl$ thin film. (b) Extracted surface roughness parameters obtained from AFM analysis. (c) Representative line-scan height profile across the film surface. (d) Cross-sectional high-resolution transmission electron microscopy (HRTEM) image of the $Mn_2CoAl/SiO_2/p$-Si heterostructure, confirming a well-defined layered structure with an ultrathin $SiO_2$ interlayer. (e) Magnetic field–dependent magnetization (*M–H*) loops measured at 10 K and 300 K. (f) Temperature-dependent magnetization and corresponding derivative (dM/dT) measured under an applied field of 500 Oe, indicating the magnetic transition temperature.

## S7. Magnetic property study

Figure S2(e) shows the magnetic field–dependent magnetization (M–H) curves of the $Mn_2CoAl$ thin film measured at 10 K and 300 K. At both temperatures, the film exhibits clear ferromagnetic hysteresis with rapid saturation at relatively low magnetic fields, indicating soft ferromagnetic behavior. At 10 K, the saturation magnetization reaches approximately 1.5-1.6 $\mu_B$/f.u., while at 300 K it decreases to about 0.7–0.8 $\mu_B$/f.u.. The experimental saturation magnetization at low temperature is reasonably close to the theoretically predicted value of 2 $\mu_B$/f.u. for inverse Heusler $Mn_2CoAl$,[8] as expected for a spin-gapless semiconductor. The strong reduction in saturation magnetization from low temperature to room temperature is primarily governed by thermal excitation of spin waves and reduced exchange alignment, while the persistence of ferromagnetism up to high temperatures confirms the thermal stability of magnetic order in $Mn_2CoAl$. The slight reduction of the experimental magnetization from the ideal theoretical value can be attributed to residual antisite disorder, finite crystallite size effects, and surface or interfacial spin disorder, which are commonly observed in sputter-deposited Heusler alloy thin films. Such imperfections can reduce the net magnetic moment by partially disrupting the ideal atomic site occupancy and magnetic exchange interactions.

Figure S2(f) displays the temperature-dependent magnetization (M-T) measured under an applied field of 500 Oe. The magnetization decreases monotonically with increasing temperature due to enhanced thermal spin fluctuations, which progressively disorder the aligned magnetic moments and reduce the net magnetization. The Curie temperature ($T_c$), estimated from the minimum of the dM/dT curve, is approximately 590 K, indicating robust ferromagnetic ordering well above room temperature.

## S8. Transport property study

Figure S3(a) illustrates the measurement geometry employed for the electrical transport studies of the $Mn_2CoAl/SiO_2/p$-Si heterostructure. The I-V characteristics exhibit strong nonlinearity and diode-like behaviour at all measured temperatures, as shown in Fig. S3(b). At finite temperatures (≥50 K), the I–V curves are nearly symmetric with respect to bias polarity, whereas a pronounced rectifying behavior emerges at 10 K. The bias symmetry observed at finite temperatures arises from thermally assisted tunneling through a broad distribution of localized states, which allows comparable carrier transmission under both bias polarities. In contrast, at 10 K, thermal activation is strongly suppressed, and transport becomes dominated by direct tunneling through energetically asymmetric barrier states dictated by the intrinsic band alignment of the $Mn_2CoAl/SiO_2/p$-Si heterostructure. This asymmetry in tunneling probability leads to the emergence of rectifying behavior at low temperature, which is distinct from conventional p–n junction rectification.

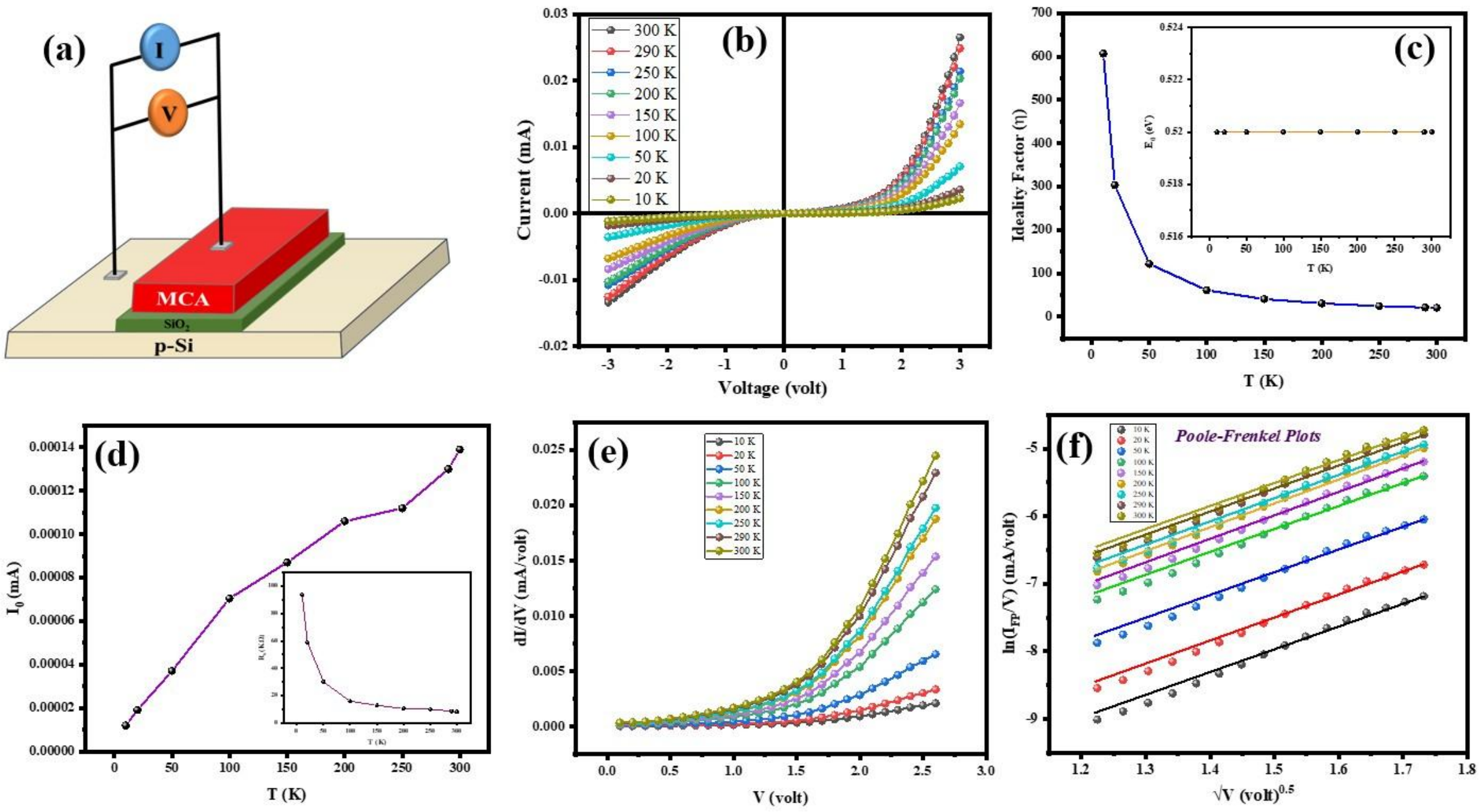

**Figure S3.** (a) Experimental setup for I–V measurements of the $Mn_2CoAl/SiO_2/p$-Si junction. (b) Temperature-dependent I–V characteristics (10–300 K) showing nonlinear, tunneling-dominated transport. (c) Ideality factor versus temperature; inset shows the temperature dependence of the tunneling constant. (d) Reverse saturation current as a function of temperature; inset displays the extracted series resistance. (e) Differential conductance (dI/dV) versus bias at different temperatures. (f) Poole–Frenkel plot [ln(I/V) vs $V^{1/2}$], confirming field-assisted transport through localized states in the $SiO_2$ barrier.

To quantitatively analyze the transport behavior, the I-V data were fitted using a modified diode equation that accounts for tunneling contributions,

$$I = I_0 \left[ \exp\left\{ \frac{qV - IR_S}{\eta \kappa_B T} \right\} - 1 \right] \tag{5}$$

where $I_0$ is the reverse saturation current, $R_s$ is the series resistance, η is the ideality factor, q is the elementary charge, $k_B$ is the Boltzmann constant, and T is the temperature. The fitted parameters are tabulated below.

**Table I.** Temperature dependence of ideality factor, tunneling constant, tunneling current and series resistance of $MCA/SiO_2/p$-Si heterostructure.

| T(K) | η | $E_0$ (eV) | $I_0$ (mA) | $R_S$ (KΩ) |
|---|---|---|---|---|
| **10** | 607.14 | 0.52 | $1.19 \times 10^{-5}$ | 93.84 |
| **20** | 303.57 | 0.52 | $1.9 \times 10^{-5}$ | 58.87 |
| **50** | 121.43 | 0.52 | $3.7 \times 10^{-5}$ | 30.14 |
| **100** | 60.7 | 0.52 | $7.04 \times 10^{-5}$ | 15.87 |
| **150** | 40.48 | 0.52 | $8.7 \times 10^{-5}$ | 12.84 |
| **200** | 30.4 | 0.52 | $1.06 \times 10^{-4}$ | 10.49 |
| **250** | 24.29 | 0.52 | $1.12 \times 10^{-4}$ | 9.97 |
| **290** | 20.94 | 0.52 | $1.3 \times 10^{-4}$ | 8.59 |
| **300** | 20.24 | 0.52 | $1.39 \times 10^{-4}$ | 8.05 |

In tunnel-dominated junctions, the physically meaningful transport energy scale is the characteristic tunneling energy $E_t = \eta\, k_B T$,[9] which reflects carrier transmission through localized states rather than thermionic emission over a barrier.

The extracted ideality factor exhibits unusually large values at low temperatures and decreases monotonically with increasing temperature, as displayed in Fig. S3(c). Such behavior should not be interpreted within the framework of conventional thermionic emission. Instead, it reflects a transport regime dominated by direct and defect-assisted tunneling through localized states in the ultrathin native $SiO_2$ barrier, where thermal activation is negligible and the current–voltage characteristics no longer follow Boltzmann statistics. In this regime, the apparent ideality factor loses its usual physical meaning and serves primarily as a fitting parameter describing the non-Boltzmann voltage dependence of tunneling current. As the temperature increases, thermally assisted tunneling becomes increasingly effective, allowing carriers to access a broader distribution of oxide trap states, which results in a progressive reduction of the apparent ideality factor.

Importantly, despite the strong temperature dependence of the extracted ideality factor, the characteristic tunneling energy scale $E_t = \eta k_B T$ remains nearly temperature independent over the entire measurement range (inset of Fig. S3(c)). In the present tunneling-dominated regime, the extracted ideality factor does not retain its conventional physical interpretation and should be regarded solely as a fitting parameter capturing the non-Boltzmann voltage dependence of defect-mediated tunneling current. The physically relevant transport parameter is the characteristic tunneling energy, which remains nearly temperature independent, confirming that transport is governed by stable interfacial and oxide-related properties rather than thermionic emission. Consistent with this interpretation, the reverse saturation current increases systematically with temperature due to enhanced thermal occupation and emission of carriers from oxide trap states (Fig. S3(d)),[9] while the series resistance decreases as a result of improved carrier mobility in *p*-Si and increased participation of thermally activated carriers (inset of Fig. S3(d)). Together, these trends provide strong evidence that charge transport across the heterojunction is governed by defect-mediated tunneling rather than ideal diode behavior.

Further insight into the transport mechanism is obtained from differential conductance (dI/dV) analysis,[10] which reveals a strongly nonlinear bias dependence over the entire temperature range (Fig. S3(e)). The absence of conductance saturation rules out space-charge-limited conduction, while the strong enhancement of conductance with increasing bias and temperature confirms electric-field-assisted carrier emission through the oxide barrier. At higher electric fields, the transport behavior follows Poole–Frenkel (PF) emission, characterized by a linear dependence of ln(I/V) on $V^{1/2}$(Fig. S3(f)).

The Poole-Frenkel emission current density is given by

$$J_{PF} = CV \exp\left\{ -\frac{q\left(\phi_B - \sqrt{qV/\pi\varepsilon_0\varepsilon_s}\right)}{\kappa_B T} \right\} \tag{6}$$

where $\phi_B$ is the barrier height for electron emission from the trap state, and $\varepsilon_s$ is the relative dielectric permittivity at high frequency, rather than the static dielectric constant. PF transport arises from electric-field-induced lowering of the Coulombic potential barrier associated with trap states in the $SiO_2$ layer, enabling field-assisted release of carriers into extended states. The dominance of PF emission is expected for ultrathin oxide barriers containing localized defects and provides an efficient high-field conduction pathway without compromising interfacial integrity.

**S9. Control Experiment with nonmagnetic Ag**

To unambiguously establish that the observed junction magnetoresistance (JMR) originates from the magnetic and spin-gapless nature of the $Mn_2CoAl$ electrode, control devices were fabricated in which the ferromagnetic $Mn_2CoAl$ layer was replaced by a nonmagnetic Ag electrode while retaining an identical device geometry and native $SiO_2$ tunnel barrier thickness. The Ag/$SiO_2$/*p*-Si junctions were prepared under identical fabrication conditions, and their electrical and magnetotransport properties were measured using the same experimental protocols.

The magnetotransport measurements reveal no measurable junction magnetoresistance within the experimental resolution over the entire temperature and magnetic-field range studied. In particular, the application of an in-plane magnetic field does not produce any systematic change in junction resistance, and no positive or negative JMR signal is detected.

The absence of JMR in the Ag-based control devices demonstrates that neither the native $SiO_2$ barrier nor the *p*-Si substrate alone can generate spin-dependent tunneling or magnetic-field-induced resistance modulation. This result conclusively shows that the large and oscillatory JMR observed in $Mn_2CoAl$/$SiO_2$/*p*-Si heterostructures arises from the presence of the magnetic, spin-polarized $Mn_2CoAl$ electrode, which enables spin-selective tunneling and the formation of an exchange-coupled, spin-polarized interfacial region in the semiconductor. The control experiment thus rules out extrinsic origins such as magnetoresistance of silicon, defect-related magnetotransport, or barrier-induced artifacts, and firmly establishes the magnetic origin of the observed JMR.

Several alternative origins of the observed junction magnetoresistance can be explicitly excluded based on the combined control experiments. First, magnetoresistance contributions from the $Mn_2CoAl$ electrode itself are ruled out by independent magnetotransport measurements performed on $Mn_2CoAl$ films deposited on electrically insulating $Al_2O_3$ substrates, where a linear, nonsaturating magnetoresistance characteristic of spin-gapless semiconductors is observed, without any junction-related resistance modulation. That means, the observed JMR does not originate from the intrinsic magnetoresistance of the $Mn_2CoAl$ film itself, but instead arises from the heterojunction ($Mn_2CoAl$/$SiO_2$/*p*-Si). Second, bulk magnetoresistance of the *p*-Si substrate cannot account for the observed effect, as it would yield a monotonic and non-oscillatory response independent of tunnel barrier thickness. Third, barrier-related magnetoresistance mechanisms, such as magnetic-field-induced barrier height modulation, wavefunction shrinkage, or magneto-impurity-assisted tunnelling, are likewise excluded, as these effects would not produce systematic sign reversals with oxide thickness. Most decisively, Ag/$SiO_2$/*p*-Si control junctions fabricated with identical geometry and native oxide thickness exhibit no measurable junction magnetoresistance within the experimental resolution, demonstrating that neither the $SiO_2$ barrier nor the *p*-Si channel alone contributes to the effect. The observed large and oscillatory junction magnetoresistance therefore uniquely originates from spin-selective tunneling enabled by the magnetic $Mn_2CoAl$ electrode.